\documentclass[12pt]{iopart}
\usepackage{graphicx}
\begin{document}
\title{Level density and $\gamma$-ray strength in $^{27,28}$Si}
\author{M.~Guttormsen\footnote[1]{To whom correspondence should be addressed (magne.guttormsen@fys.uio.no)}, E.~Melby, J.~Rekstad, and S.~Siem}
\address{Department of Physics, University of Oslo, N-0316 Oslo, Norway}
\author{A.~Schiller}
\address{Lawrence Livermore National Laboratory, L-414, 7000 East Avenue,
Livermore CA-94551, USA}
\author{T.~L{\"o}nnroth}
\address{Department of Physics, {\AA}bo Akademi, FIN-20500 Turku, Finland}
\author{A.~Voinov}
\address{Frank Laboratory of Neutron Physics, Joint Institute of Nuclear
Research, 141980 Dubna, Moscow reg., Russia}

\begin{abstract}
A method to extract simultaneously level densities and $\gamma$-ray
transmission coefficients has for the first time been tested on light nuclei
utilizing the $^{28}$Si($^3$He,$\alpha\gamma)^{27}$Si and
$^{28}$Si($^3$He,$^3$He'$\gamma)^{28}$Si reactions. The extracted level
densities for $^{27}$Si and $^{28}$Si are consistent with the level densities
obtained by counting known levels in the respective nuclei. The extracted
$\gamma$-ray strength in $^{28}$Si agrees well with the known $\gamma$-decay
properties of this nucleus. Typical nuclear temperatures are found to be
$T\sim 2.4$~MeV at around 7~MeV excitation energy. The entropy gap between
nuclei with mass number $A$ and $A\pm 1$ is measured to be
$\delta S\sim 1.0\ k_B$, which indicates an energy spacing between
single-particle orbitals comparable with typical nuclear temperatures.
\end{abstract}

\submitto{\JPG}


\maketitle

\section{Introduction}

The Oslo Cyclotron group has developed a method to extract first-generation
(primary) $\gamma$-ray spectra at various initial excitation energies
\cite{gutt0}. From the set of primary spectra, nuclear level densities and
$\gamma$-ray strength functions can be extracted \cite{hend1,schi0}. These two
functions reveal essential nuclear structure information such as shell gaps,
pair correlations, and thermal and electromagnetic properties. In the last
couple of years, the Oslo group has demonstrated several fruitful applications
of the method \cite{melb0,schi1,gutt1,gutt2,schi2,gutt3,melb1,voin1,siem1}.

So far, the method has been tested on rare earth nuclei, having a rather
uniform and high single-particle level density. These properties are expected
to be important for the foundation of the method. The two crucial steps in the
extraction procedure are:\ (i) the subtraction technique for obtaining the
primary $\gamma$-ray spectra, and (ii) the Brink-Axel hypothesis telling that
nuclear resonances with approximately equal properties can be built on all
states. Thus, the most important requirements are that the spin and parity
distributions should be similar for all excitation energy bins and that the
nucleus should thermalize at each step within a given $\gamma$ cascade.

Although not finally validated, the method has been tested and found successful
for deformed rare-earth nuclei
\cite{melb0,schi1,gutt1,gutt2,schi2,gutt3,melb1,voin1} and even for the weakly
deformed $^{148,149}$Sm nuclei \cite{siem1}. However, in cases where the
statistical properties are less favourable, as for lighter nuclei and/or for
nuclei in the vicinity of closed shells, the foundation of the method is more
doubtful. Since both $^{27}$Si and $^{28}$Si of the present study are
open-shell systems in the middle of the $sd$ shell, only the statistical
question remains here. 

From a statistical point of view it is important 
that for light nuclei having low level density one can expect
a strong influence of Porter-Thomas fluctuations on 
$\gamma$-transition intensities of primary $\gamma$ spectra and hence, 
on extracted level densities and
$\gamma$-ray strength functions. This means that the
main assumptions of the Oslo method can fail, namely that
the applicability of the Axel-Brink hypothesis can loose its meaning
because of large $\gamma$-intensity fluctuations and insufficient averaging over
nuclear levels due to a low level density. From this point of view it is
interesting to establish the limit of the applicability of the Oslo method 
by investigating light nuclei.

In this work, we have tested the extraction procedure on the light $^{27}$Si
and $^{28}$Si systems, applying the ($^3$He,$\alpha$) pick-up reaction and the
($^3$He,$^3$He') inelastic scattering reaction, respectively. Since both the
level density and the $\gamma$-decay rates are known in these nuclei, a
quantitative judgment of the applicability of the procedure is feasible.
Section 2 describes the experimental methods and techniques, and in sections 3
and 4 the level density and $\gamma$-ray strength are discussed. Finally,
concluding remarks are given in section 5.

\section{Experimental method and techniques}

The experiment was carried out at the Oslo Cyclotron Laboratory with a 45~MeV
$^3$He beam. The self-supporting $^{28}$Si target was isotopically enriched to
$100\,\%$ and had a thickness of 3~mg/cm$^2$. The reactions employed in the
extraction procedure were $^{28}$Si($^3$He,$\alpha\gamma)^{27}$Si and
$^{28}$Si$(^3$He,$^3$He'$\gamma)^{28}$Si.

The charged particles and $\gamma$ rays were recorded with the detector array
CACTUS, which contains eight particle telescopes, 27~NaI $\gamma$-ray detectors
and two additional Ge detectors monitoring the spin distribution and the
selectivity of the reactions. Each telescope is placed at an angle of
45$^{\circ}$ relative to the beam axis, and comprises one Si front and one
Si(Li) back detector with thicknesses of 140 and 3000~$\mu$m, respectively. The
NaI $\gamma$-detector array surrounding the target and particle detectors has a
resolution of $\sim 6\,\%$ at $E_\gamma=1$~MeV and a total efficiency of
$\sim 15\,\%$.

\begin{figure}
\includegraphics[totalheight=17.5cm,angle=0,bb=0 80 350 730]{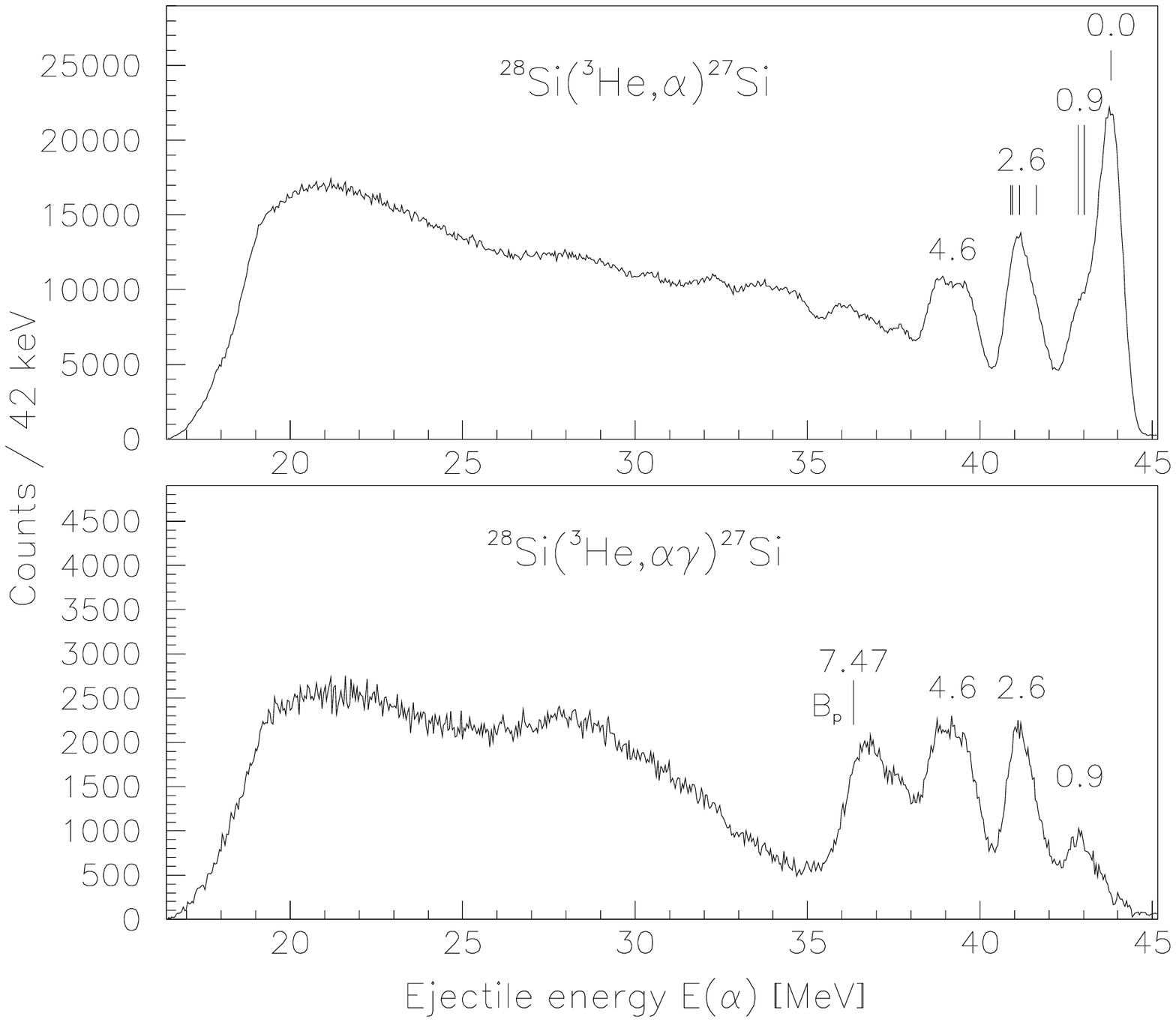}
\caption{$\alpha$-particles from the $^{28}$Si$(^3$He,$\alpha)$ reaction
measured at 45$^{\circ}$ with respect to the beam axis and with a beam energy
of 45~MeV. The lower panel shows the same spectrum, but in coincidence with one
or more $\gamma$ rays detected in the NaI detectors. Some known levels and the
proton binding energy are indicated in MeV and by vertical lines.}
\label{fig:pp14he}
\end{figure}

\begin{figure}
\includegraphics[totalheight=17.5cm,angle=0,bb=0 80 350 730]{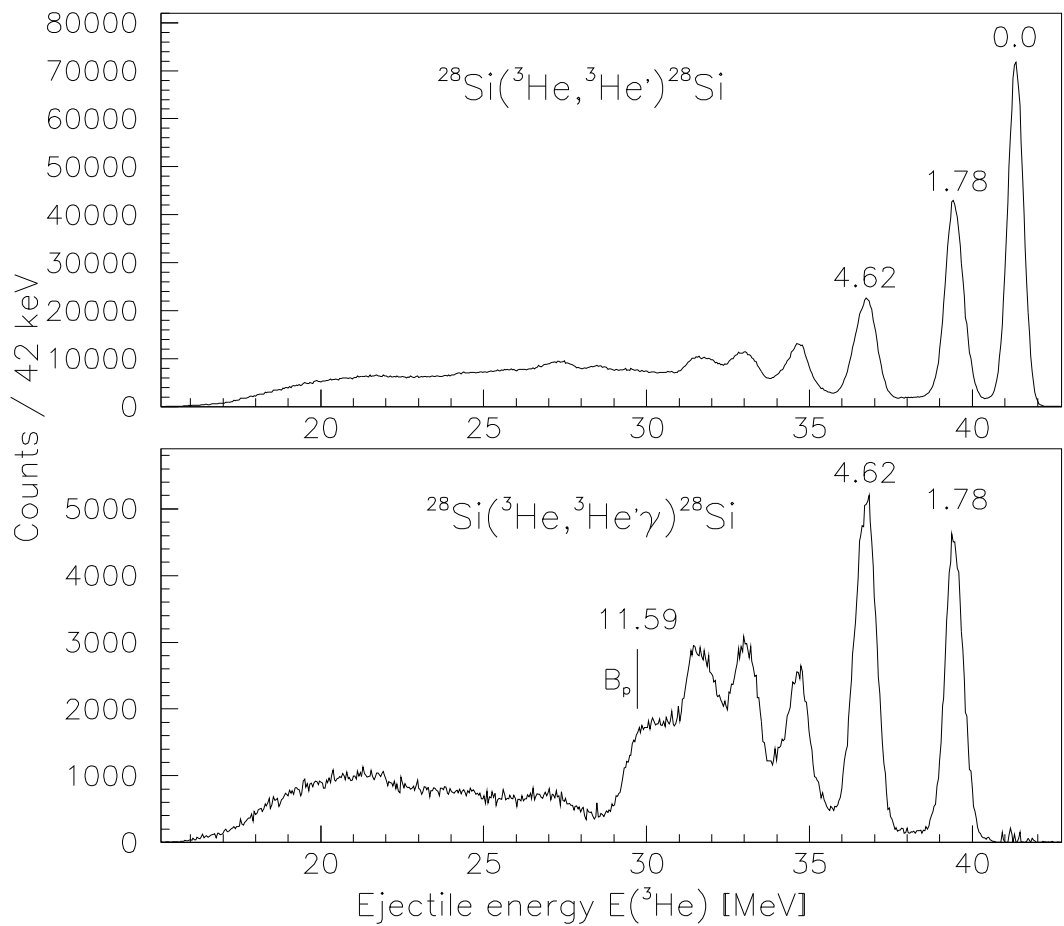}
\caption{Same as figure 1, but $^3$He particles from the
$^{28}$Si($^3$He,$^3$He') reaction.}
\label{fig:pp13he}
\end{figure}

In figures \ref{fig:pp14he} and \ref{fig:pp13he}, the single and coincidence
ejectile spectra are shown for the reactions
$^{28}$Si$(^3$He,$\alpha)^{27}$Si and $^{28}$Si$(^3$He,$^3$He'$)^{28}$Si,
respectively. The ground state in $^{27}$Si is located at higher ejectile
energy due to a positive $Q$ value of 3.40~MeV in the $(^3$He,$\alpha)$
reaction. In $^{27}$Si and $^{28}$Si, pronounced peak structures are seen up to
5 and 10~MeV excitation energy, respectively. The coincident spectra show a
drop in yield at the proton binding energies $B_p$, since the residual nucleus
with one proton less (aluminium) is then populated at low excitation energy with
a corresponding low $\gamma$-ray multiplicity. Our extraction method is
therefore only applicable up to the excitation energy of $E\sim B_p$. The spin 
window of the present pick-up and elastic scattering reactions is 
concentrated at $I\sim 2-4 \hbar$.

\begin{figure}
\includegraphics[totalheight=17.5cm,angle=0,bb=0 80 350 730]{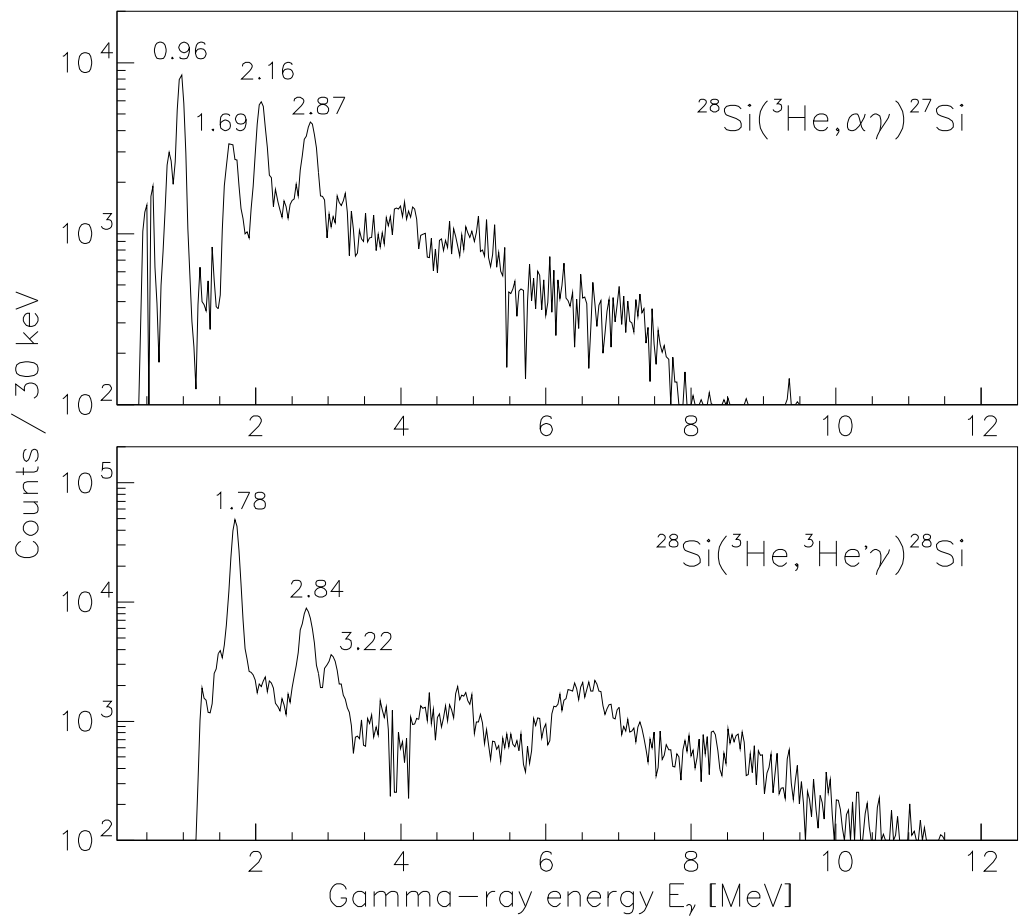}
\caption{Unfolded NaI $\gamma$-ray spectra from $^{27}$Si and $^{28}$Si. The
strongest $\gamma$ lines are indicated by energies in MeV.}
\label{fig:ug34helog}
\end{figure}

In order to determine the true $\gamma$-energy distribution, the $\gamma$
spectra are corrected for the response of the NaI detectors with the unfolding
procedure of \cite{gutt4}. In addition, random coincidences are subtracted
from the $\gamma$ spectra. The resulting unfolded NaI $\gamma$ spectra are
shown in figure \ref{fig:ug34helog}. The spectra reveal strong $\gamma$ lines,
which are identified as transitions between low-lying states.

The set of unfolded $\gamma$ spectra are organized in a $(E,E_\gamma)$ matrix,
where the initial excitation energies $E$ of $^{27}$Si and $^{28}$Si are
determined by means of reaction kinematics, utilizing the energy of the
ejectile. This matrix comprises the $\gamma$-energy distribution of the total
$\gamma$ cascade. The primary-$\gamma$ matrix can now be found according to the
subtraction technique of \cite{gutt0}.

\begin{figure}
\includegraphics[totalheight=17.5cm,angle=0,bb=0 80 350 730]{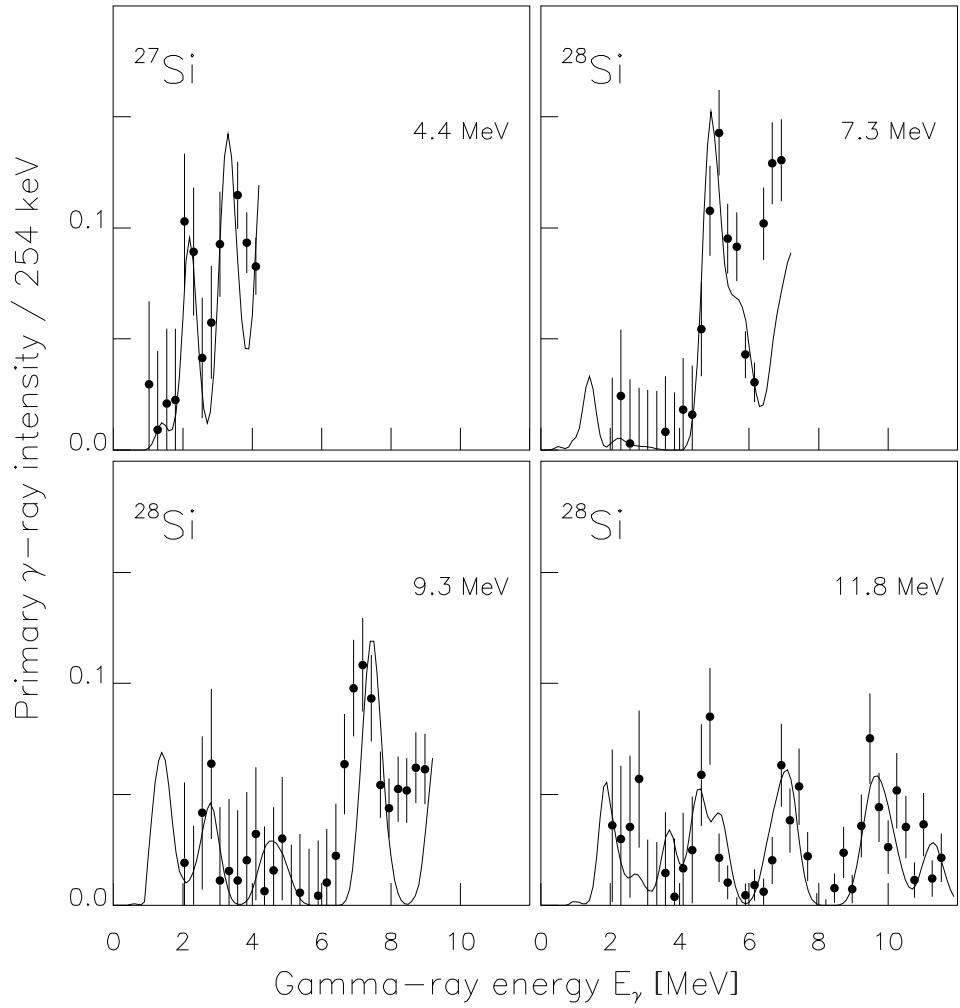}
\caption{Comparison between primary $\gamma$-ray spectra extracted with our
method (data points with error bars) and the spectra (solid lines) constructed
from the known \protect\cite{ENSDF} $\gamma$-decay branching of levels around
excitation energies indicated for each panel. The spectra are normalized to
unity.}
\label{fig:firstgenexp}
\end{figure}

As indicated in the introduction, the primary $\gamma$-ray procedure is based
on the assumption that the decay properties of the particular reaction-selected
distribution of states within each energy bin are independent on whether the
respective ensembles of states are directly populated through the nuclear
reaction or by $\gamma$ decay from higher-lying states. In the present case,
the $\gamma$-ray multiplicity is low and reaches $M_\gamma\sim 2$ around 10~MeV
excitation energy. Thus, generally less than half of the counts in the total
$\gamma$ spectra have to be subtracted in the procedure, giving quite reliable
primary $\gamma$-ray spectra. This is demonstrated in
figure \ref{fig:firstgenexp}, where we compare our results to known
$\gamma$-decay branches from various excitation-energy regions \cite{ENSDF}. 
Here, in the reconstruction of primary $\gamma$ spectra from known $\gamma$
transitions, we have assumed equal population of the initial levels within each 
energy bin. Taking into account that probably not all levels and branchings are known, the agreement is very satisfactory. It should be noted that possible Porter-Thomas fluctuations will reveal themselves in exactly the same way in the literature data and the extracted primary $\gamma$ spectra. Thus, agreement between the two spectra of figure \ref{fig:firstgenexp} should not be surprising from a statistical point of view.

The idea is now to find two functions, the level density $\rho(E)$ and the
$\gamma$-energy-dependent function ${\mathcal T}(E_\gamma)$, that, multiplied with each
other, reproduce the set of primary $\gamma$-ray spectra. The ${\mathcal T}(E_\gamma)$
function is identified as the so-called $\gamma$-ray transmission coefficient.
According to the Brink-Axel hypothesis \cite{brink,axel}, the primary
$\gamma$-ray spectra can be factorized by
\begin{equation}
P(E,E_\gamma)\propto\rho(E-E_\gamma){\mathcal T}(E_\gamma),
\label{eq:ab}
\end{equation}
where $P(E,E_\gamma)$ is fitted to the primary $\gamma$-ray matrix by a least
$\chi^2$ fit \cite{schi0}. In the factorization procedure, we use primary
spectra from the initial excitation-energy regions of $E=3.6$--7.6~MeV and
$E=4.2$--12.1~MeV for $^{27}$Si and $^{28}$Si, respectively. The corresponding
$\gamma$-energy regions were chosen to be $E_\gamma=1.1$--7.6~MeV and
$E_\gamma=2.2$--12.1~MeV.

\begin{figure}
\includegraphics[totalheight=17.5cm,angle=0,bb=0 80 350 730]{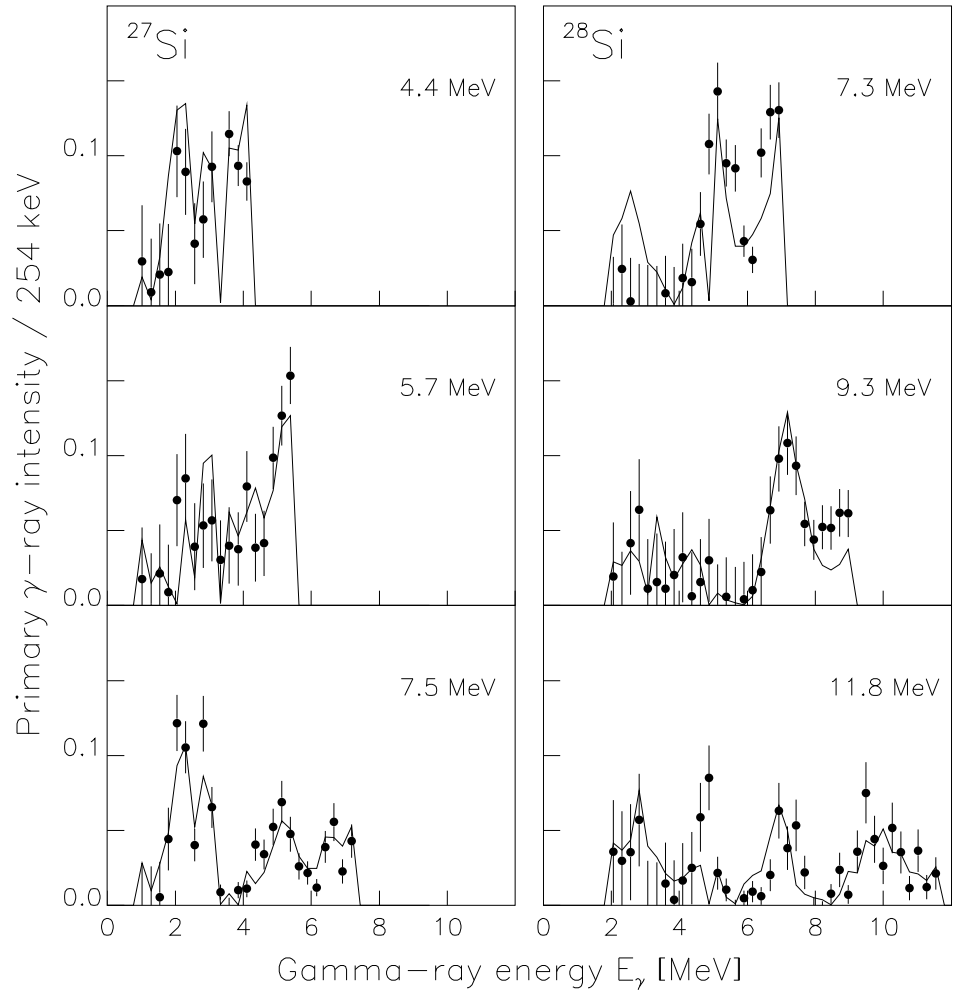}
\caption{Comparison between experimental primary $\gamma$-ray spectra (data
points with error bars) and the spectral distributions (solid lines) calculated
from the extracted level density $\rho(E)$ and the $\gamma$-ray transmission
coefficient ${\cal T}(E_\gamma)$. The initial excitation-energy bins are indicated
in each panel. The spectra are normalized to unity.}
\label{fig:fgteo}
\end{figure}

In figure \ref{fig:fgteo}, the best fits (solid lines) to the experimental $P$
are shown for the $^{27}$Si and $^{28}$Si nuclei. The experimental spectra exhibit strong peak structures 
that are in good correlation in spite of the fact that
they have been obtained at different excitation energies. The
correlation indicates that the peak structures are mainly due to the $\gamma$-ray transmission 
coefficient ${\cal T}(E_{\gamma})$ which is assumed
to be independent of the excitation energies of initial and final states. This
is also supported by the fact that the
highly structured experimental spectra are seen to be fairly well described by the
factorization in equation (\ref{eq:ab}). 
The expected Porter-Thomas fluctuations can be estimated from the deviations
of the experimental points from
the solid line in figure \ref{fig:fgteo}, which demonstrate that the fluctuations
are of minor importance for the extracted level densities and $\gamma$-ray
transmission coefficients.

In our extraction procedure, we assume that the $\gamma$-ray transmission 
coefficient ${\cal T}(E_{\gamma})$ is independent of the excitation energy 
(or temperature). Although no strict violation of this assumption has been determined
experimentally, theoretical predictions show that ${\cal T}(E_{\gamma})$ may depend on 
the thermal properties of the final state. In order to estimate the influence of 
such effects, we have 
applied the temperature dependent model of Kadmenski{\u{\i}}, Markushev and 
Furman \cite{KMF} for electric dipole transitions. The model describes 
a $\gamma$-ray strength function based on the low energy tail of the 
giant electric dipole resonance (GEDR), where we use GEDR parameters 
from \cite{TD98}. To evaluate the temperature at 
a certain excitation energy, we use $T=\sqrt{U/a}$, where $U$ and $a$ are the 
intrinsic excitation energy and the level density parameter, respectively. 
These parameters are determined 
according to von Egidy {\it et al.} \cite{Egidy}. In our experiment, we measure 
$\gamma$ spectra from a certain initial excitation energy $E$, and the 
temperature $T$ is determined by the excitation energy of the final state
$E-E_{\gamma}$. Since we use spectra from several $E$ for a given $E_{\gamma}$, also
the final excitation energy will change accordingly, giving different temperatures
for different $\gamma$ energies. For the tests, we used the KMF model including
this variation in temperature with $\gamma$ energies, and compared with a calculation where the temperature is kept constant. Here,
we find for $^{28}$Si that the $\mathcal{T}_{\rm KMF}(E_{\gamma})$ values deviate within 20 \% for $E_{\gamma} > 3$ MeV. 
However, for lower $E_{\gamma}$ the $\mathcal{T}_{\rm KMF}(E_{\gamma})$ 
value with variable temperature increases and gains a factor of two at $E_{\gamma} = 2$ MeV compared to the value of $\mathcal{T}_{\rm KMF}(E_{\gamma})$ taken
at a constant temperature. The varying temperature is associated with a standard deviation of $\sim 0.4$ MeV, which will reduced this difference somewhat.  
Since the predicted temperature dependence is highly uncertain, if existing at all in real silicon nuclei, we omit to include this uncertainty in the experimental error bars.

The multiplicative functions $\rho(E)$ and ${\mathcal T}(E_\gamma)$ of equation (\ref{eq:ab}) are not uniquely determined. Actually, one
may construct an infinite number of other functions \cite{schi0}, which give
identical fits to the data of figure \ref{fig:fgteo} by
\begin{eqnarray}
\tilde{\rho}(E-E_\gamma)&=&A\exp[\alpha(E-E_\gamma)]\,\rho(E-E_\gamma),
\label{eq:array1}\\
\tilde{\mathcal T}(E_\gamma)&=&B\exp(\alpha E_\gamma){\mathcal T}(E_\gamma).
\label{eq:array2}
\end{eqnarray}
Consequently, neither the slope nor the absolute value of the two functions can
be obtained through the fitting procedure. Thus, the free parameters $A$, $B$ and
$\alpha$ have to be determined from other information to give the best physical
solution.

\section{Level density}

The $^{27}$Si and $^{28}$Si nuclei have been thoroughly studied during the last
decades, and the level schemes are probably fairly complete up to several MeV
excitation energy. By smoothing the known level density taken from the database
of \cite{ENSDF} with the experimental energy resolution, we can adjust the
parameters $A$ and $\alpha$ of our experimental $\rho$ to fit the level density
to the number of known levels. The adjustments were performed at $E=0.9$ and
5.0~MeV in $^{27}$Si and at $E=1.7$ and 7.0~MeV in $^{28}$Si. These anchor
points are well determined and separated in excitation energy, however, also
other data points could as well be used.

\begin{figure}
\includegraphics[totalheight=17.5cm,angle=0,bb=0 80 350 730]{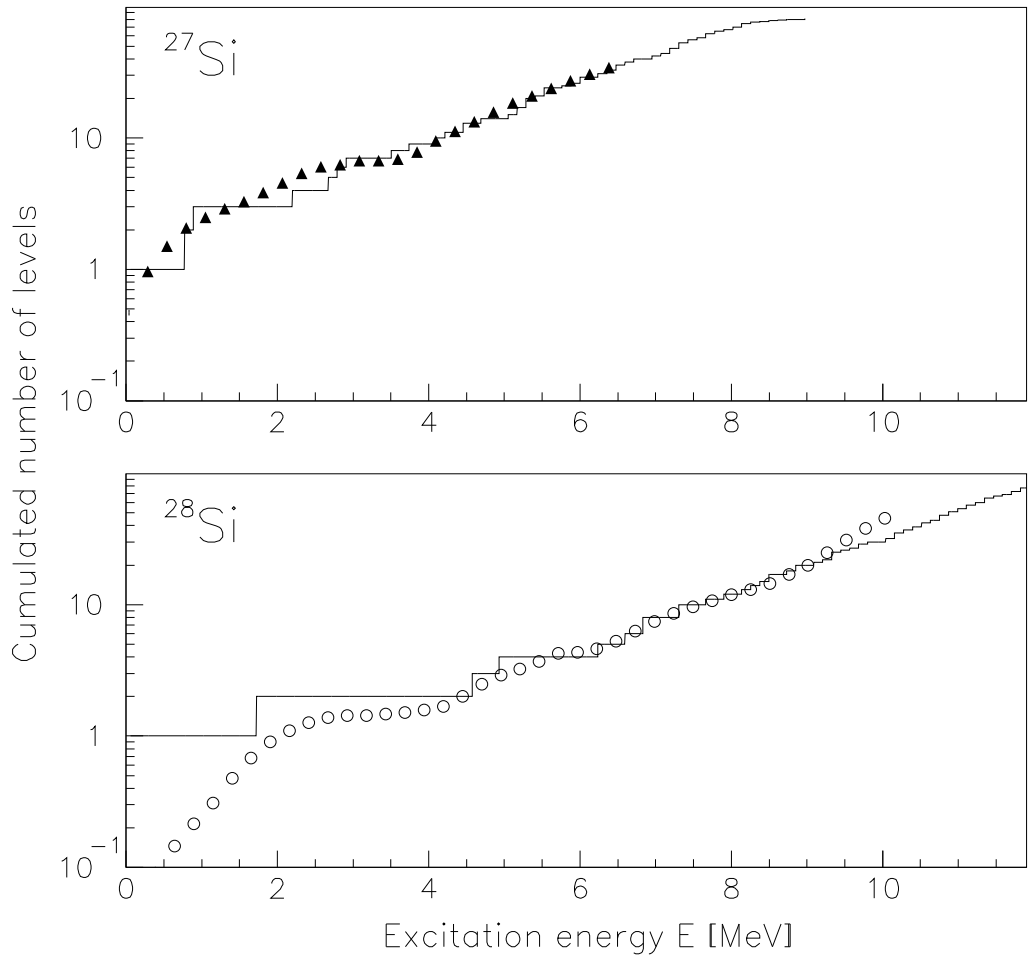}
\caption{Cumulated number of levels with energy up to $E$ in $^{27}$Si and $^{28}$Si. 
The extracted cumulative numbers (data points) are compared to values based on 
known levels (histogram). The extracted data for $^{28}$Si at lower 
energies is underestimated due to the weak direct $\gamma$ feeding
of the ground state from $E = 4.2-12.1$ MeV of excitation energies.}
\label{fig:acc34he}
\end{figure}

Figure \ref{fig:acc34he} compares cumulated number of levels
based on our data (data points) and
known levels (histogram) taken from \cite{ENSDF}. The overall
agreement is gratifying\footnote[7]{Apparently, our data seem to be wrong adjusted to the anchor point at $E=1.7$ MeV in $^{28}$Si, but 
the plot shows cumulated numbers of levels and
not level densities.}. Local differences can be caused by violation of the
Axel-Brink hypotheses for light nuclei, where low level density and large
fluctuations of $\gamma$ intensity is observed. For example, in $^{28}$Si, the
level density around the ground state is strongly underestimated in the experiment,
because very few transitions decay directly to the $I^\pi=0^+$ ground state.
This property is less pronounced in $^{27}$Si, having a ground state assignment
of $I^\pi=5/2^+$. Since these nuclei have been well studied, we are not able
to extract much more information on the level density than what is already
known. However, our data give a small indication that not all levels in
$^{28}$Si have yet been observed above 9~MeV excitation energy.

Very few explicit calculations of level densities exist. This would be needed,
since a simple Fermi-gas approximation based on the level-density parameter $a$
and the pairing-gap parameter $\Delta$ neglects residual interactions. In the
calculations of Ormand \rm\cite{orm97}, the level densities of
$^{24}$Mg and $^{32}$S were studied in a Monte-Carlo Shell-Model approach
within the $sd$ shell. It is interesting that the level density in $^{24}$Mg
behaves very much like the nuclei studied here. The fact that there are only
one hundred levels per MeV at excitation energies of 8--12~MeV has
implications. It indicates that if specific states are selected by some
reaction, e.g., $\alpha$-cluster states with large parentage of $\pi^2\nu^2$
configurations, the number of states of this kind necessarily must be
significantly lower. However, at this stage, it is not possible to give
quantitative statements of the number of, e.g., $\alpha$-cluster states in this
region.

In general, the most successful way to create additional states in atomic
nuclei is to break up $J=0$ nucleon Cooper pairs. Each broken
pair increases the level density by a factor of 10-20, which is much higher than obtained from
other modes of excitation as, e.g., rotation or vibration. Recently
\cite{gutt3,gutt5,schi3}, a thermodynamic model was developed which could describe
level densities for mid-shell nuclei in the mass regions $A=58$, 106, 162, and
234. The model is based on the canonical ensemble theory, where equilibrium is
obtained at a certain given temperature $T$.

The basic idea of the model \cite{gutt3,gutt5,schi3} is the assumption of a reservoir
of proton and neutron pairs. The pairs may be broken such that unpaired
nucleons are thermally scattered into an infinite, equidistant,
doubly-degenerated single-particle level scheme. In addition, rotational and
vibrational modes may be thermally excited. The total partition function is
written as a product of proton $(Z_\pi)$, neutron $(Z_\nu)$, rotation
$(Z_{\mathrm{rot}})$, and vibration $(Z_{\mathrm{vib}})$ partition functions.
Then, from the Helmholtz free energy
\begin{equation}
F(T)=-T\ln\left(Z_\pi Z_\nu Z_{\mathrm{rot}}Z_{\mathrm{vib}}\right),
\end{equation}
thermodynamical quantities as entropy, average excitation energy, and heat
capacity can be calculated as
\begin{eqnarray}
S(T)&=&-\left(\frac{\partial F}{\partial T}\right)_V\\
\langle E(T)\rangle&=&F+ST\\
C_V(T)&=&\left(\frac{\partial\langle E\rangle}{\partial T}\right)_V,
\end{eqnarray}
where the Boltzmann constant is set to unity $(k_B=1)$ and the temperature $T$
is measured in MeV. The level density is calculated applying the saddle-point
approximation \cite{NA97}
\begin{equation}
\rho(\langle E\rangle)=\frac{\exp (S)}{T\sqrt{2\pi C_V}}.
\end{equation}

\begin{figure}
\includegraphics[totalheight=17.5cm,angle=0,bb=0 80 350 730]{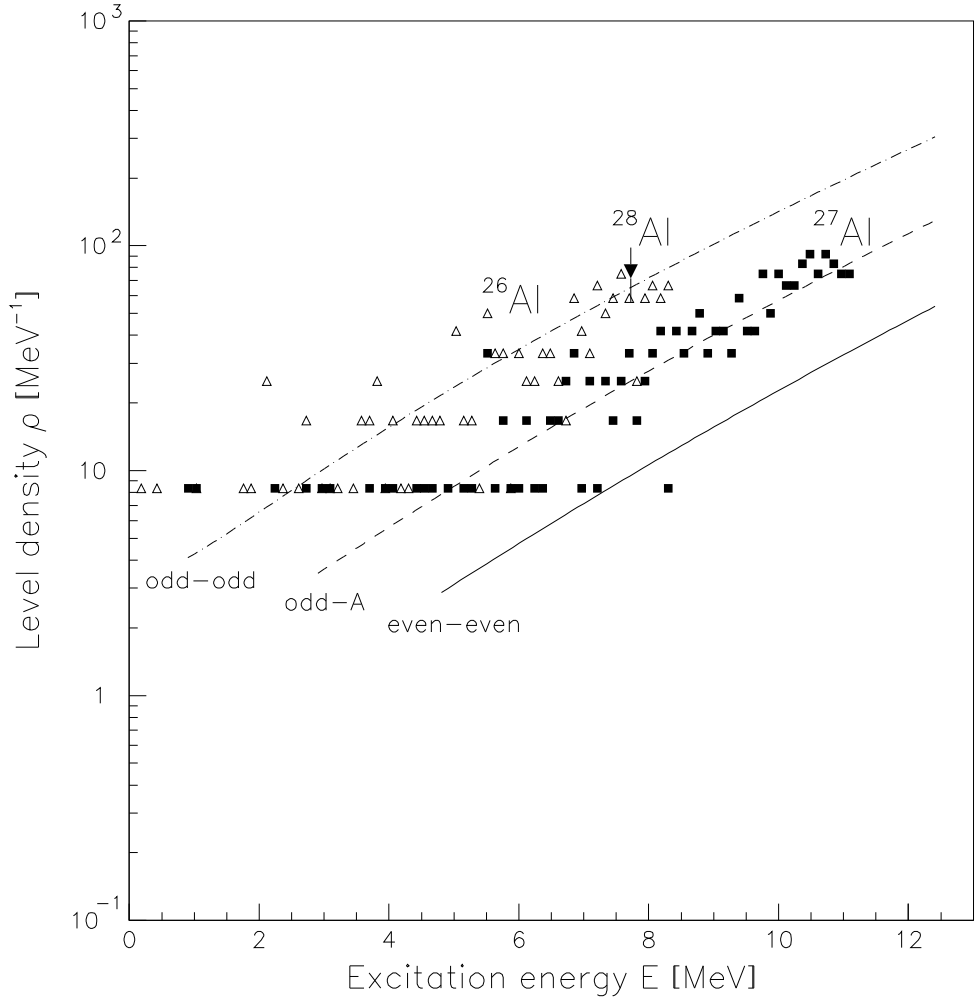}
\caption{Level density in $^{26}$Al (open triangles) and $^{27}$Al (filled
squares) determined by counting the number of known levels \protect\cite{ENSDF}
within excitation energy bins of 120~keV. One data point (filled triangle with error
bar) is based on neutron-resonance spacings in $^{28}$Al
\protect\cite{gutt3,Iljinov}. The model calculations (curves) are performed
with the same parameter set for the even-even, odd-$A$ and odd-odd systems:
$\varepsilon=2.0$~MeV, $\Delta=2.3$~MeV, $r=0.56$, $A_{\mathrm{rot}}=0.3$~MeV,
$\hbar\omega_{\mathrm{vib}}=5.0$~MeV, and with seven proton and seven neutron
pairs in the reservoir.}
\label{fig:aluteo}
\end{figure}

The energy spacing between the single-particle orbitals $\varepsilon$ is the
most critical parameter of the model. In order to determine $\varepsilon$, we
investigate the well-studied $^{26}$Al and $^{27}$Al isotopes, where the level
density is presumably known up to high excitation energy from the counting of
experimentally observed levels \cite{ENSDF}. Figure \ref{fig:aluteo} shows
these experimental data together with a level-density point evaluated for
$^{28}$Al \cite{gutt3} from the average neutron-resonance spacing
\cite{Iljinov} at the neutron binding energy $B_n$. The slopes of the level
densities are well reproduced with $\varepsilon=2.0$~MeV, which is consistent
with the energy gaps in the Nilsson single-particle level scheme. The
difference in level density between $^{26}$Al and $^{27}$Al depends strongly on
the pairing-gap energy and is well described by a gap parameter of
$\Delta=12A^{-1/2}$~MeV$\ =2.3$~MeV, using $A=28$\footnote[2]{It
is difficult to determine an appropriate pairing gap parameter in this mass region. 
If one calculates pairing gap 
parameters $\Delta_p$ and $\Delta_p$ from nuclear mass differences \cite{Bohr}, the various systems give strongly scattered values. This mainly reflects the variation in the distance between the Fermi level and neighbouring single particle states \cite{Dobaczewski}, which could in principle be as high as $\varepsilon \sim 2.0$~MeV. Since this distance is schematically included in our model, we adopt here the "pure" pairing contribution of approximately $\Delta=12A^{-1/2}$~MeV, which represents an average value in this mass region.}.

The pairing-attenuation
factor, defined as the ratio between the amount of energy needed to break one
nucleon pair compared to the previous broken pair, is set to $r=0.56$ (for
details see \cite{schi3}). For simplicity, we have used the same values
of $\varepsilon$, $\Delta$, and $r$ for protons and neutrons. Furthermore, all
calculations include a reservoir of seven proton and seven neutron pairs. The
collective parameters are taken from experimental data on $^{28}$Si. The
rotational parameter $A_{\mathrm{rot}}=E/2(2+1)=0.3$~MeV is calculated from the
first excited state $(I^\pi=2^+)$, and the vibrational energy
$\hbar\omega_{\mathrm{vib}}=5.0$~MeV is set equal to the excitation energy of
the first non-rotational state $(I^\pi=0^+)$.

\begin{figure}
\includegraphics[totalheight=17.5cm,angle=0,bb=0 80 350 730]{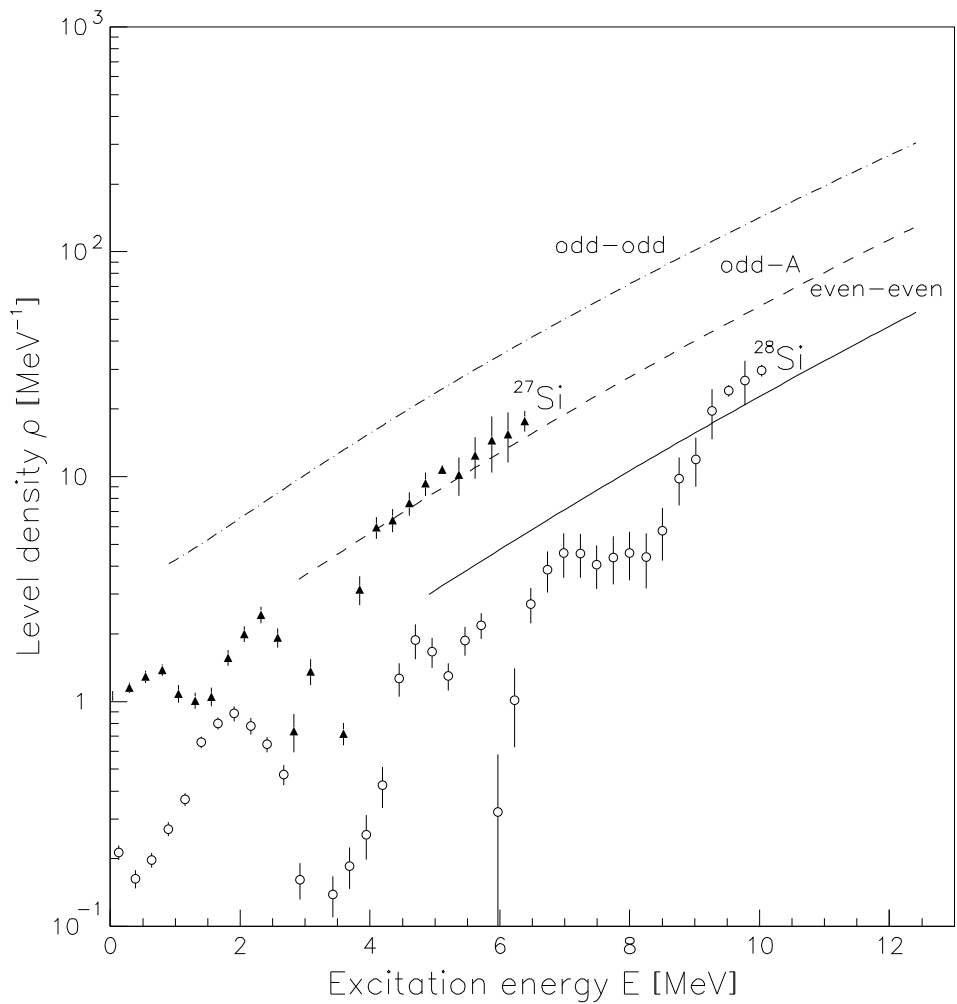}
\caption{Comparison between level densities extracted from primary $\gamma$-ray
spectra in $^{27}$Si and $^{28}$Si (data points) and model predictions
(curves), see text of figure 7.}
\label{fig:expteo}
\end{figure}

The very same parameter set is used in the calculation for $^{27}$Si and
$^{28}$Si in figure \ref{fig:expteo}. The calculation describes well the
experimental data of $^{27}$Si for excitation energies above $E\sim 4$~MeV. For
$^{28}$Si, the more statistical and smooth part of the level density takes
place first above $E\sim 9$~MeV, making it difficult to judge the   
agreement between model and experiment in this case.

\begin{figure}
\includegraphics[totalheight=17.5cm,angle=0,bb=0 80 350 730]{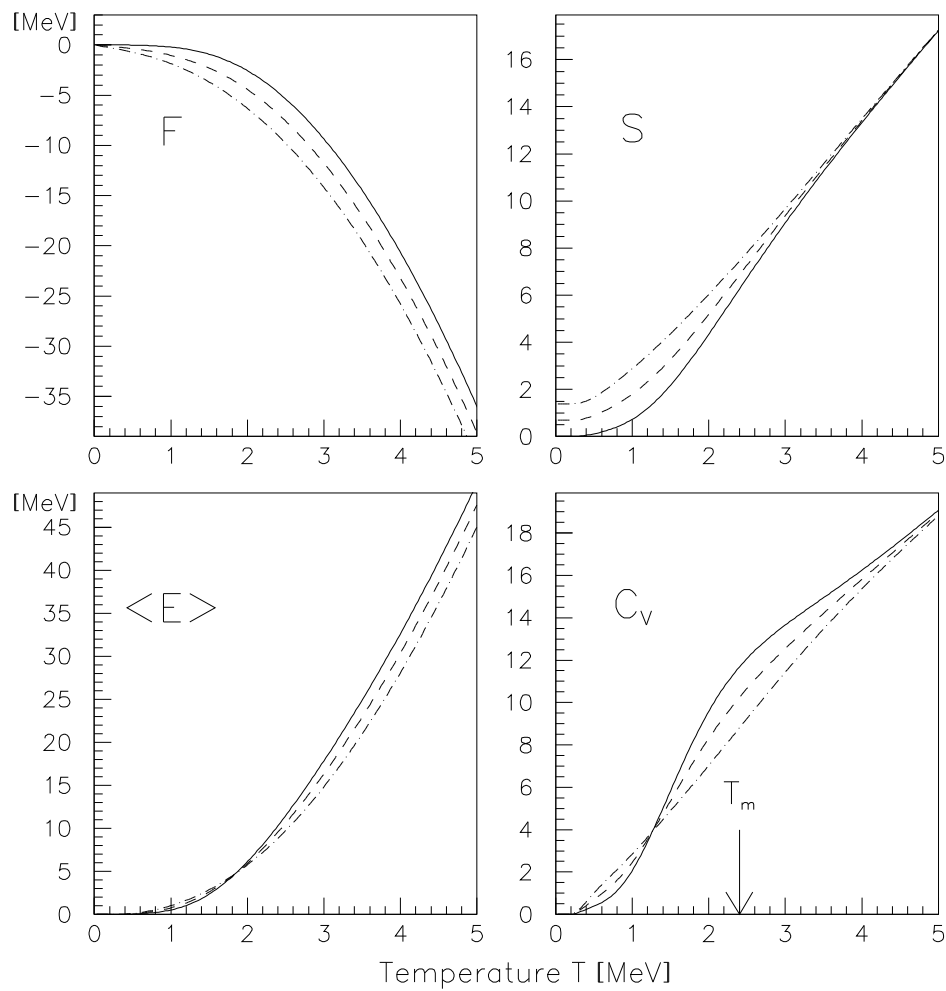}
\caption{Model calculations for nuclei around $^{28}$Si showing even-even
(solid line), odd-$A$ (dashed line) and odd-odd (dash-dotted line) systems. The
four panels show the free energy $F$, the entropy $S$, the average excitation
energy $\langle E\rangle$, and the heat capacity $C_V$ as function of
temperature $T$. The arrow at $T_m=2.4$~MeV indicates the local maximum of
$C_V$, where the pair breaking process is strong. The parameters are given in the
text of figure 7.}
\label{fig:sfec}
\end{figure}

By comparing the level densities between even-even ($^{28}$Si), odd-$A$
($^{27}$Al or $^{27}$Si), and odd-odd ($^{26}$Al) systems\footnote[3]{The proton
and neutron parameters are identical and give no difference between odd-even
and even-odd systems. We therefore denote either of these two systems as the
odd-$A$ system.}, we may extract interesting information. In
figure \ref{fig:sfec}, the free energy $F$, entropy $S$, average excitation
energy $\langle E\rangle$, and heat capacity $C_V$ are shown as functions of
temperature for the three systems. A fruitful quantity is the entropy
difference $\delta S$ between systems with $A$ and $A\pm 1$. At low
temperatures, this quantity is approximately extensive (additive) and
represents the single-particle entropy associated with the valence particle (or
hole) \cite{gutt3}. In the upper right panel, we find that the nucleon carries
a single-particle entropy of $\delta S\sim 1.0$ at $T\sim 1.5$~MeV.

The single-particle entropy can also be deduced from the experimental data of
aluminium (and possibly also from silicon). The high-energy data points of
figure \ref{fig:aluteo} reveal that the level density of $^{26}$Al is $\sim 2.7$
times higher than for $^{27}$Al. Remembering that the level density and entropy
are connected by
\begin{equation}
S=\ln\rho+{\rm constant},
\label{eq:lnrho}
\end{equation}
we deduce also here $\delta S\sim 1.0$.

The statistical definition of temperature in the microcanonical ensemble is
given by
\begin{equation}
T=\left(\frac{\partial S}{\partial E}\right)^{-1}_V,
\label{eq:t}
\end{equation}
giving typically $T\sim 2.4$~MeV for the data points of $^{27}$Al in the
excitation-energy region of 7--11~MeV, see figure \ref{fig:aluteo}. Since the
excitation-energy shift between the level densities of figure \ref{fig:aluteo}
amounts approximately to the pairing-gap parameter $\Delta$, we may express the
slope of $\ln(\rho)$ as $\delta S/\Delta$ or, according to
equations (\ref{eq:lnrho}) and (\ref{eq:t}), as $T^{-1}$, giving
\begin{equation}
\delta S=\Delta/T,
\end{equation}
which again confirms $\delta S=2.3/2.4\sim 1.0$.

\begin{figure}
\includegraphics[totalheight=17.5cm,angle=0,bb=0 80 350 730]{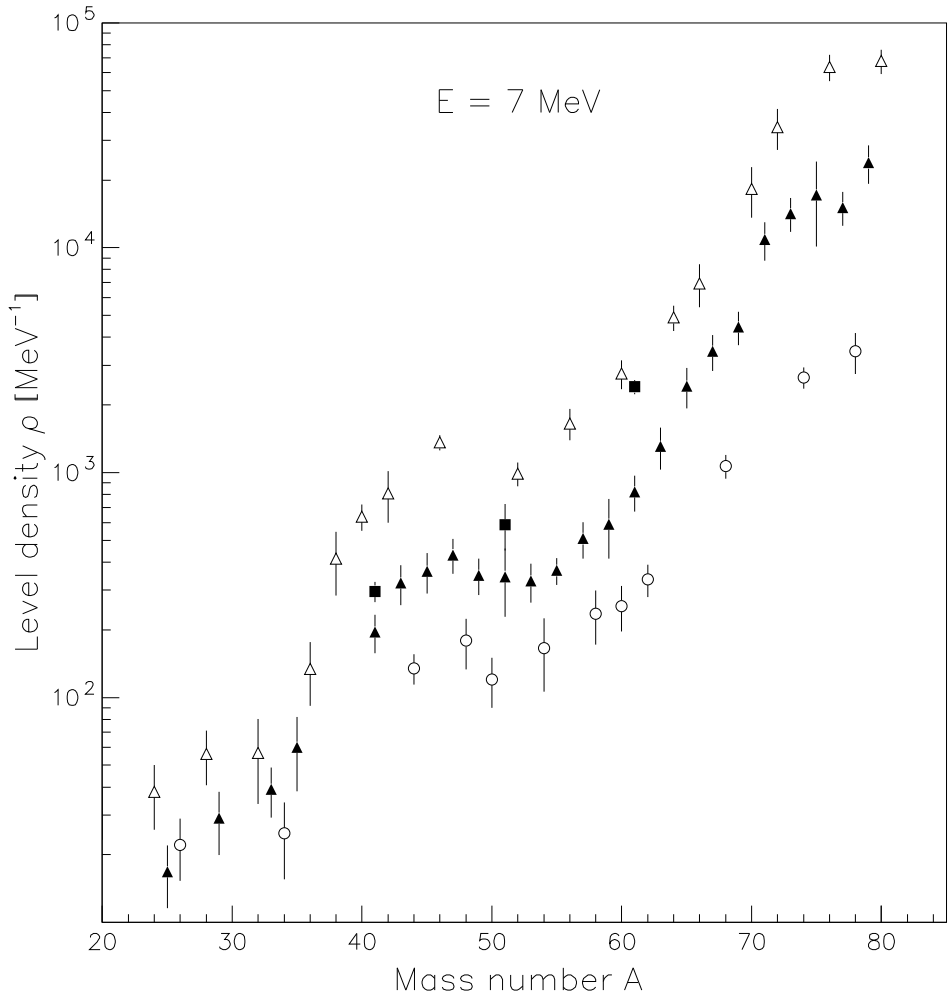}
\caption{Level densities as function of mass number at 7~MeV excitation energy.
The data are plotted for odd-odd (open triangles), odd-even (filled triangles),
even-odd (filled squares), and even-even (open circles) nuclei. The data are
extracted from experimental values, as described in \protect\cite{gutt3}.}
\label{fig:counta7}
\end{figure}

The extracted single-particle entropy is lower than observed in heavier
mid-shell nuclei where typically $\delta S\sim 1.7$ \cite{gutt3}. The recent
findings are supported by the systematics of figure \ref{fig:counta7}, where the
experimentally deduced level densities at $E=7$~MeV are displayed for various
nuclear systems. 
These data points are evaluated as a linear interpolation of 
$\ln \rho(E)$ between two anchor points: (i) the level density based on known levels at low excitation energy and (ii) the level density estimated from neutron resonance energy spacings at the neutron binding energy \cite{gutt3}. Figure \ref{fig:counta7} 
shows larger error bars and more scattered data for mass numbers below $A\sim 40$, which mainly reflects that the first anchor point is strongly influenced by local shell effects. 
However, in spite of local shell effects below $A\sim 40$, it is 
clear that the level densities drop
one order of magnitude, and the level density gaps between 
various systems are severely reduced. From our model calculations, this change
is due to a decrease in the ratio between the temperature $T$ and the
single-particle energy spacing $\varepsilon$. In the mass region studied here,
the temperature and spacing are approximately equal, i.e., $T\sim\varepsilon$,
and the valence nucleon is thermally smeared over the ground state and the
first excited single-particle level, giving $\delta S=\ln 2=0.7$, only. In the
rare earth region, we have typically $T\sim 3\varepsilon$ and the valence
nucleon is thermally spread over several single-particle levels, giving higher
entropy.

The heat-capacity curves of figure \ref{fig:sfec} show S-shaped forms, in
particular for the even-even system. The local maximum of this shape appears at
$T_m\sim 2.4$~MeV, corresponding to an excitation energy of
$\langle E\rangle=10$~MeV. At this point, the excitation energy increases
strongly with temperature, indicating a substantial breaking of nuclear Cooper pairs. It is
tempting to connect this overshoot in heat capacity with a pairing-phase
transition. In order to test this, we have studied the distribution of zeros of
the partition function in the complex temperature plane, as prescribed in
\cite{schi3,CS00}. In our calculations, we find that the zeros move
away from the real, inverse-temperature axis with increasing inverse
temperature. This property is inconsistent with a pairing-phase transition in
the thermodynamical limit of $A\rightarrow\infty$, in accordance with the
generalized Ehrenfest definition of a phase transition \cite{Ehren}. A
similar conclusion has also been drawn for the Fe mass region \cite{schi3}.

\section{Gamma-ray strength}

\begin{figure}
\includegraphics[totalheight=17.5cm,angle=0,bb=0 80 350 730]{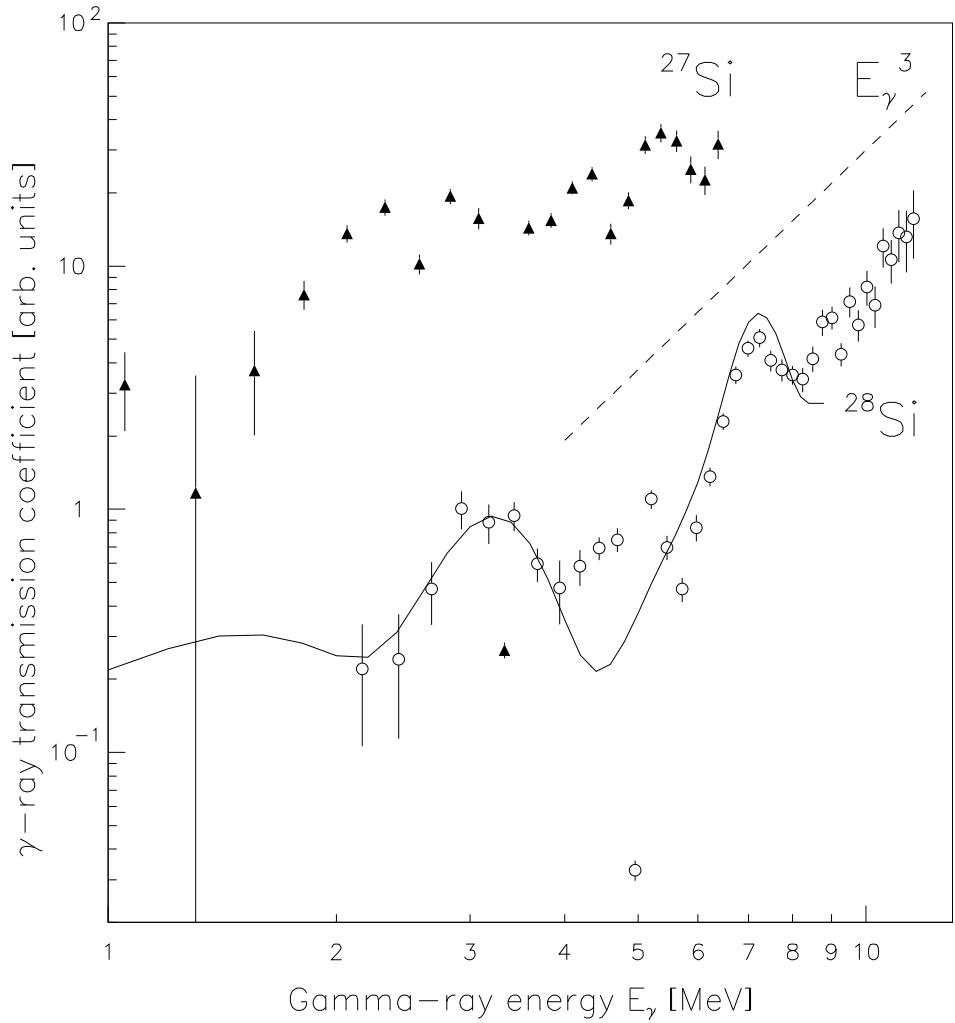}
\caption{Extracted $\gamma$-ray transmission coefficient ${\cal T}(E_\gamma)$ in
$^{27}$Si and $^{28}$Si (not normalized). The solid line is determined (see
text) from $\gamma$ transitions depopulating levels in $^{28}$Si with known
lifetimes \protect\cite{ENSDF}. The slopes of the transmission coefficients
follow very roughly a $\sim E_\gamma^3$ dependency (dashed line).}
\label{fig:strength}
\end{figure}

The $\gamma$-ray energy-dependent function ${\cal T}(E_\gamma)$ contains information
on the average $\gamma$-decay probability. In figure \ref{fig:strength}, the
non-normalized ${\cal T}(E_\gamma)$ of equation (\ref{eq:array2}) is shown with the
$\alpha$ parameter determined in section 3. The two extreme data points at
$E_\gamma=3.2$ and 5.0~MeV for $^{27}$Si and $^{28}$Si, respectively, are
probably due to a singularity in the extraction method, giving too small error
bars.

It has been shown \cite{Blatt,Barth} that the regularity of the $\gamma$-decay
strength in the continuum is determined by the $\gamma$-ray strength
function\footnote[5]{Also called the radiative strength function in literature.}
\begin{equation}
f_{XL}=\Gamma_{XL}^i(E_{\gamma})/(E_{\gamma} ^{2L+1}D^i),
\label{eq:fxl}
\end{equation}
where $E_\gamma$ is the transition energy, $X$ denotes the electric or magnetic
character, $L$ is the multipolarity, $\Gamma_{XL}^i(E_\gamma)$ is the
partial radiative width, and $D^i$ is the level spacing of states with spins
and parities equal to those of the initial state $i$. The main part of
high-lying levels is known to decay by $E1$ and $M1$ $\gamma$ transitions.
Thus, one could in principle extract $f_{E1}+f_{M1}$ from the present
experiment in the same way as has been done for several rare-earth nuclei
\cite{melb1,voin1,siem1}. However, the concept of an average radiative width
is difficult to adopt for these light nuclei, since there is no experimental
information on the mean values of the level spacing $D^i$; there are less
than one level per MeV with presumably the same spin and parity.

Even though the absolute strength of $f$ is uncertain, its functional
dependence on $E_\gamma$ can be studied. The shape of the measured $\gamma$-ray
transmission coefficient is given by
\begin{equation}
{\cal T}(E_\gamma)=2\pi\,\sum_{\lambda=1,2,\cdots}[f_{E\lambda}(E_\gamma)+f_{M\lambda}(E_\gamma)]E_\gamma^{2\lambda+1}.
\end{equation}
If the transitions are mainly governed by
single-particle dipole transitions, an $E_\gamma^3$ shape would be expected according
to the Weisskopf estimate, where $f_{E1}$ and $f_{M1}$ take constant values. 
The functional form is displayed in
figure \ref{fig:strength}. Both $^{27}$Si and $^{28}$Si show a rather flat shape
of ${\cal T}$ for $E_\gamma<4$--5~MeV. However, for $E_\gamma>4$--5~MeV, $^{28}$Si
follows roughly the $E_\gamma^3$ dependence, indicating that the transitions
are rather of single-particle than collective origin as found for rare-earth
nuclei \cite{melb1,voin1,siem1}. Also, the large spread in lifetimes
\cite{ENSDF} of the high-lying levels supports this interpretation. 

In $^{28}$Si, levels and their lifetimes and $\gamma$-decay branching ratios
are known up to 9.6 MeV of excitation energy from discrete $\gamma$
spectroscopy \cite{ENSDF}. Thus, we may define the decay intensity in terms
of the partial ($\Gamma_{ij}$) and total ($\Gamma_i$) decay widths by
$I_{ij}(E_\gamma)=\Gamma_{ij}/\Gamma_i$, where the $\gamma$-transition energy is
determined by the excitation energies of the initial ($i$) and final ($j$)
states by $E_\gamma=E_i-E_j$. For this case, the $\gamma$-ray transmission
coefficient reads
\begin{equation}
{\cal T}_{ij}(E_\gamma)=2\pi\hbar I_{ij}(E_\gamma)/\tau_i\,D^i,
\end{equation}
where $\tau_i$ is the lifetime of level $i$. Here, we roughly estimate $D^i$
 from the total level density 
using the spin distribution of \cite{Gilbert}. The constructed
$\gamma$-ray transmission coefficient is then smoothed as a function of
$E_\gamma$ with the experimental resolution. Despite the fact that not all
levels and decay branches are known, figure \ref{fig:strength} shows that the
smoothed $\gamma$-ray transmission coefficient compares surprisingly well
with the extracted data points. This indicates that the method may work even
in cases where no well-behaving smooth $\gamma$-ray strength function is
present.

\section{Conclusions}

A simultaneous extraction of level densities and $\gamma$-ray transmission
coefficients in $^{27,28}$Si has been performed. Comparison to literature data
gives excellent agreement within the experimental resolution. This study
demonstrates that the extraction method also works for light nuclei, despite
the fact that thermalization might seem questionable. The concept of a
$\gamma$-ray strength function in this mass region is not very useful. The
measured $\gamma$-energy dependence of approximately $E_\gamma^3$ indicates
that the $\gamma$ decay is governed by single-particle dipole transitions.

A simple thermodynamical model reproduces the high excitation-energy region of
the experimental data. For $^{27,28}$Si and $^{26,27}$Al, we
conclude that a nuclear temperature of $T\sim 2.4$~MeV and a pairing gap 
parameter of
$\Delta\sim 2.3$~MeV describes the level densities at high excitation energies. 
Both experiment and model indicate that the valence
nucleon in this mass region carries an entropy of $\delta S\sim 1.0$. This is
close to the expected value of 0.7 when the thermal nucleonic excitation is
smeared over the ground state and only one excited single-particle level. The
pair breaking process is strongly smeared out in temperature (and excitation
energy) and a phase transition in the sense of the Ehrenfest definition is
doubtful.

The present work encourages also similar studies in the mass region
$A=30$--100. Here, nuclear level densities are used in the determination of 
nuclear reaction cross sections from Hauser-Feshbach type of calculations. 
These cross sections, as well as the $\gamma$-ray strength functions, are 
important inputs in large network calculations of stellar evolution
\cite{Rauscher}, as well as in the simulations of accelerator-driven 
transmutation of nuclear waste.

\section*{Acknowledgements}
Financial support from the Norwegian Research Council (NFR) is gratefully
acknowledged. Part of this work was performed under the auspices of the U.S.
Department of Energy by the University of California, Lawrence Livermore
National Laboratory under Contract No.\ W-7405-ENG-48. T.L. warmly acknowledges
the financial support from the Magnus Ehrnrooth Foundation in Helsinki. A.V.
would like to thank the staff of the University of Oslo for their warm
hospitality while this manuscript was prepared and acknowledges support from a
NATO Science Fellowship under project number 150027/432 given by the Norwegian
Research Council (NFR).

\clearpage

\clearpage

\clearpage

\clearpage

\clearpage

\clearpage

\clearpage

\clearpage

\clearpage

\clearpage

\clearpage

\end{document}